# Electron self-injection for the acceleration in laser-pulse-wakes in the presence of a 'strong' external magnetic field

A. Zhidkov[1,2,3], T. Hosokai[1,2], S. Masuda[1,2], Y. Oishi[3], T. Fujii[3], R. Kodama[1]

[1]*Photon Pioneers Center, Osaka University, 2-8, Yamadaoka, Suita, Osaka 565-0871, Japan*
[2]*Japan Science and Technology Agency (JST), CREST, 2-8, Yamadaoka, Suita, Osaka 565-0871, Japan*
[3]*Central Research Institute of Electric Power Industry, 2-6-1 Nagasaka, Yokosuka, Kanagawa 240-0196, Japan*

**Abstract**

An external static magnetic field with its strength $B\sim10T$ may result in the laser wake wave-breaking upon changing the electron motion in the vicinity of maximal density ramp of a wave period. This, as shown by numerical simulations, can change the resonance character of the electron self-injection in the laser wake-field; a total charge loaded in the acceleration phase of laser pulse wake can be controlled by a proper choice of the magnetic field strength.

**I. Introduction**

In the last decade, many research groups have reported on the direct observations of mono-energetic electron bunches with a low energy spread after irradiating gas jets by ~10-100 TW femtosecond laser pulses [1-5]. This has boosted the interest in the laser wake field acceleration (LWFA) as a possible practical way to produce compact, high energy electron accelerators. However while in some experiments mono-energetic beams with hundreds of MeV energy have been observed [1,2,5] in the others, conducted almost at the same conditions, the maximal electron energy was not high, ~100 MeV [6]. Moreover in all experiments, fluctuations of the total charge and maximal bunch energy were far higher than the fluctuation of laser pulse parameters. The pre-pulse effects [3], which may be strong in higher density plasmas, cannot explain such a fluctuation in lower density gas targets used for the GeV-level electron acceleration. In all those experiments, accelerated electrons appeared due to their self-injection, which is a mechanism of electrons loading into the acceleration phase of the laser wake, and, therefore, the source of the instabilities should be sought in this injection process.

In Ref. [7] the resonance character of electron self-injection has been demonstrated by means of multidimensional particle-in-cell simulation. The waist of a laser pulse propagating in plasma is not constant; it changes periodically. The amplitude and frequency of such oscillations depend on plasma transverse gradients and pulse intensity, if relativistic self-focusing is dominant. For example in a plasma channel and a lower laser power, the waist oscillation frequency approximately equals $\Omega \sim c\omega_{pl}^m/(\omega R_0)$, where $\omega_{pl}^m$ and $\omega$ are the maximal plasma frequency and laser frequency, $R_0$ is the radius of plasma channel. If $\Omega$ becomes closer to the plasma frequency in the reference frame moving with the laser pulse in plasma equal $\tilde{\omega}_{pl} = \omega_{pl}/2\gamma$ [with the relativistic factor determined by the pulse group velocity $v_{gr} = c\sqrt{1 - N_e/N_{cr}}$ which gives: $\tilde{\omega}_{pl} = \omega N_e(0)/2N_{cr}$], a parametric resonance may occur. According to Ref. [7], a small variation in plasma parameters results in a large change of charge loading and, therefore, in the maximal energy of accelerated electrons, becoming a source of instability. It is difficult to make a proper control of beam parameters, especially in non full-optical acceleration schemes.

Effects of external magnetic fields on LWFA (a case of ultra-strong $B>100T$ magnetic field has been discussed in Ref. [8]) have yet to be well studied. In Ref [3] the use of magnetic field has been proposed for modification of pre-pulse effects including formation of plasma channels [9] and, hence, an indirect control of the electron self-injection. However, a relatively high magnetic field $B\sim 10T$ directed along the propagation of laser light can be used for the control of the electron injection as well. In the moving reference frame the Larmor frequency, $\Omega_L = eB/mc$, may approach the plasma frequency for $B > B_{sh} = mc\omega_{pl}^2/2\omega$ [for $N_e \sim 10^{18}$ cm$^{-3}$ and $\omega \sim 2.3 \times 10^{15}$ s$^{-1}$: $B_{sh} \sim 5$T] and, therefore, may result in the self-injection at any ratio of $\Omega/\tilde{\omega}_{pl}$ (see Ref.[7]). The necessary strength of magnetic field can be produced by the exploding foil technique [10] in which the magnetic field up to 20T is generated in 100 ns. Since the laser wake field acceleration runs usually in few picoseconds, the proper matching is warranted within the strength from 0T to 20T.

In the present paper, we numerically study effects of a 'strong', ~10 T, external magnetic field on the electron self-injection and acceleration by the laser wake field in a pre-ionized plasma channel as well as in uniform pre-ionized plasma by means of three dimensional particle-in-cell simulations.

## II. Laser waist oscillations and density modulation

Propagation of a laser pulse in plasma over many Rayleigh lengths, which is necessary for LWFA, is usually sustained by the pulse refraction on density or/and relativistic electron mass gradients [7]. The intensity of a guided pulse is not constant. To demonstrate this we consider a pulse focused at the edge of a plasma channel with the density profile as following: $N_e(r) = N_e(0) + N_e(R_0)r^2/R_0^2 = N_e(0)[1 + \delta r^2]$, where $R_0$ is the channel radius, $N_e(0)$ is the electron density at the channel axis; $\delta$ is a parameter. Neglecting by the electron evacuation produced by the ponderomotive force of laser pulses, we can use Maxwell equation is the form:

$$\left[\frac{\partial^2}{\partial x^2} + \nabla_\perp^2 - \frac{1}{c^2}\frac{\partial^2}{\partial t^2}\right]\vec{E} = \frac{\omega_{p0}^2}{c^2\bar{\gamma}}\left(1+\delta r^2\right)\vec{E} \quad (1)$$

In the parabolic approximation it can be solved analytically assuming $\bar{\gamma} \approx \bar{\gamma}_0 + ur^2$ and neglecting the small terms $\sim r^4$; the right term is zero at $x \leq 0$. A parabolic solution is sought in the form: $\vec{E} = e^{i\omega t - ikx + A(x) - D(x)r^2/w_0^2}$; $D(0) = 1$; $A(0) = \ln|E(0,0,0)|$, where $w_0$ is the waist size in the focus point, $\omega_{p0}$ is the plasma frequency at the axis. Finally in dimensionless units, $\tilde{x} = x/L_R$, $\tilde{r} = r/w_0$; $L_R = kw_0^2/2$, one can get the following equations:

$$iA' = -D; -iD' + \alpha^2 = D^2;$$

with $\alpha^2 = w_0^2 \omega_{p0}^2 (\delta + u)/c^2 \bar{\gamma}_0$. The real part of a solution for $D$ gives a form of waist evolution:

$$D(\tilde{x}) = \tilde{\alpha}\frac{2\alpha + i(1-\alpha^2)\sin(2\alpha\tilde{x})}{1+\alpha^2 - (1-\alpha^2)\cos(2\alpha\tilde{x})}, \quad (2)$$

[At $x<0$ the solution is the upper limit of $D$ for $\alpha\to 0$]. The amplitude ratio in the absolute units is $\Gamma = \alpha^{-2} = c^2\bar{\gamma}/(\omega_{p0}^2\delta w_0^4)$ and the oscillation frequency is $\Omega_W = 2\alpha c/L_R = [4N_e(R_0)c^2/N_{cr}R_0^2\bar{\gamma}_0]^{1/2}$. A typical value of $\Gamma$ is in the range $10^{-3}$-$10^{-2}$ depending on the pulse waist, while $\Omega_W$ does not depend on the pulse waist.

In the case of relativistic intensities of laser pulses, the solution cannot be found in an analytical form. In this case we may use the approach proposed in Ref [11,7].

$$\frac{d^2U}{dx^2} \approx \frac{2}{kL_R}\frac{1}{U^3} - \frac{N_e(R_0)}{N_{cr}}\left(\frac{2w_0}{R_0}\right)^2\frac{U}{\bar{\gamma}_0} - \frac{N_e(0)}{N_{cr}}\frac{a_0^2}{4(\bar{\gamma}_0 U)^3} \quad (3)$$

where $L_R=kw_0^2/2$ is the Rayleigh length, $a_0$ is the normalized laser field, $U=D(x)/w_0$, $x$ is normalized by the laser waist in the focus point, $w_0$, $k=\omega/c$, and $\gamma = (1+a_0^2/2)^{1/2}$. The oscillation period can be estimated for a small amplitude oscillation from Eq. (1) assuming $U=U_0+\Delta$ and $\Delta<<U_0$: $\Omega_w = \sqrt{4N_e(R_0)c^2/N_{cr}\bar{\gamma}_0 R_0^2}$. The frequency from Eq.(3) equals the frequency from the paraxial approximation, Eq. (2), the relativistic effect occurs here in the reduction of oscillation amplitude. One has to remember that we neglect the electron evacuation by the ponderomotive force resulting in the density modulation in the front of the laser pulse. This non-linear effect also results in the waist oscillation.

The waist oscillates transversely. In the moving reference frame its value is also given by Eq. (3), while the plasma frequency in that reference frame equals:

$$\Omega_{pl} = ckN_e(0)/2N_{cr}$$

Since the $N_e(R_0)$ and $N_e(0)$ are independent parameters, the waist period can be numerically close to the plasma oscillation period that may result in the parametric resonance with the charge loading sensitive to plasma parameters as shown in Ref. [7].

An external magnetic field with its strength $B_{ext}\sim$1-100 T will hardly result in the high-intensity-laser-pulse dynamics. In Eq.(1) the external magnetic field gives the term order of $VH_{ext}/cE_{las} \ll 1$, where $V$ is the electrons mean velocity, $E_{las}$ is the laser field strength. However, a relatively strong magnetic field may affect the parametric resonance. In the moving reference frame and an external magnetic field directed along the pulse propagation, plasma electron will oscillate with the Larmor frequency:

$$\Omega_H = eB_{ext}/mc\gamma',$$

where $\gamma'$ is the electron relativistic factor, which is different from that determined by the pulse group velocity. In this reference frame, where electrons are in rest, the general solution for the electron motion in crossing transverse $E\cos(\Omega_{pl}t)$ and longitudinal $B$ fields is as the following:

$$x + iy = iA + De^{-i\Omega_H t} + 0.5ia\left[e^{i\Omega_{pl}t}/(\Omega_H + \Omega_{pl}) - e^{-i\Omega_{pl}t}/(\Omega_H - \Omega_{pl})\right] \quad (4)$$

where $A$ and $D$ are real constants, $a=eE_{pl}/m\Omega_{pl}$ with $E_{pl}$ the plasma field strength. The Larmor frequency becomes closer to the plasma frequency at $B\sim$10T for the conditions under consideration. In this case the drift acceleration in the crossing magnetic and plasma electric fields, according to Eq. (4), may become as strong as that by the plasma field. In the vicinity of the maximal density ramp moving with the group velocity in the laboratory frame electron fluid motion is changed in the presence of magnetic field. Therefore, a magnetic field, similar to waist oscillations, may result in the amplitude and phase of consecutive plasma wave periods, and even abolish the resonance when the Larmor oscillation becomes dominant. To reveal the effect of external magnetic fields we performed 3D particle-in-cell simulations.

**III. Results of 3D particle-in-cell simulations and discussion**

It is quite clear that the effect of magnetic field can be studied only numerically. We use three dimensional fully relativistic particle-in-cell simulations exploiting the moving window technique [7] to study electron self-injection and their further acceleration by laser pulse wake. The maximal size of simulation box was 120x85x85 $\mu m^3$; the spatial resolution was $\lambda/16$ in the direction of pulse propagation and $\lambda/8$ in the transverse direction, where $\lambda$ is the laser pulse wavelength; 8 particles per cell have been set. Laser pulses move from right to left in figures.

The plasma parameters were set as in Ref. [7]. However, a higher intensity, $I=5x10^{19}$ W/cm$^2$, and electron density cases were also examined. In these cases the external magnetic field does not result in the propagation of laser pulses, and all effects occurring are a result of wake interaction with the magnetic field. We also consider the laser pulse propagation in uniform plasma in the presence of a strong external magnetic field. The magnetic field strength was varied from 0T to 100T.

Dynamics of laser pulse field in a plasma channel and in uniform plasma is illustrated by Fig.1 and Fig.2. For the parameters chosen, there is no strong self-focusing; such regimes are more appropriate for studying the resonance electron self-injection [7]. In the case of plasma channel there is no essential change in the pulse waist that is in accordance with Eq. (2),(3). In contrast, in uniform plasma the oscillation amplitude is high ~30%; the strong diffraction loss is observed at the minimal pulse waist- the maximal focusing. Finally, the total depletion in uniform plasma becomes higher than that in a plasma channel. This results in the maximal energy of accelerated. The radiation from accelerated electrons is seen at rear of the laser pulses and is shown by arrows in the figures.

We do not observe any essential difference in the laser pulse propagation with and without magnetic field. However, the magnetic field may affect the plasma parameters through the breakdown by the laser pre-pulse [3]. Here we suppose that all possible effects of laser pre-pulses are already included in plasma parameters.

In Fig. 3, spatial distributions of accelerated electrons in a plasma channel with $N_e(0)=1.3 \times 10^{18}$ cm$^{-3}$, $N_{emax}=4 \times 10^{18}$ cm$^{-3}$, the channel diameter $D=60$ μm, and the laser pulse intensity $I=10^{19}$ W/cm$^2$, pulse duration $\tau=40$ fs and $w_0=12$ μm is given for the magnetic field strengths $B$=0, 10, 20, and 100T after 4 mm of propagation. The strong dependency on the magnetic field strength is clearly seen. The difference in the spatial distribution is coursed by a difference in the injection mechanisms with and without the magnetic field. While the parametric resonance results in straightforward self-injection [7], the magnetic-field-induced self-injection to occur requires the transverse electron motion. As a result in the magnetic field, the electrons acquire energy gains in the transverse direction as well which appear in the form of bow waves [12]. Only part of these electron are trapped and accelerated in straight forwardly. All distribution are almost symmetrical. At B=100T, the beam emittance seems to become impractical, however the geometrical angle of the cone is only 40 μm/4mm~$10^{-3}$!

According to Fig.4, numerically, the energy distribution of accelerated electrons does not depend on the magnetic field essentially. The maximal energy of accelerated electrons is about 400 MeV for all cases. However the charge of accelerated electrons is different. In Fig.5 we present the charges of accelerated electrons in different energy range: 50-100,100-200,200-300,300-400 MeV. One can see the strong dependency of the charges on the magnetic field. This is also seen in Fig.4 as density of dots representing particles. For the calculation parameters the resonance condition are close to $B=10T$ that one can see from the calculations. The change of channel parameter usually results in the charge distribution. In Fig.6 we give the comparison between the calculation done in Ref.[7] with $B=0T$ and the present calculation for $B=10T$. The effect of magnetic injection increases the total charge of accelerated electrons only slightly at the optimal conditions for the self-injection at the parametric resonance. Its effect becomes much stronger out of resonance. That means in LWFA with use of capillary plasma and magnetic field with strength ~10T the injected charge will be always order of 100 pC and weakly fluctuating with the plasma channel and laser focusing parameters then in experiments [1,2,5].

We also observe an effect of magnetic field at $B$~10T for higher density plasma which cannot be explain by the resonance as in Eq. (2). We compare the electron self-injection and acceleration in a plasma channel and uniform plasma at higher density $N_e(0)=10^{19}$ cm$^{-3}$ and a higher laser intensity. The electron density distribution in the laser wake for this case is given in Fig. 7, 8 . Fig. 7a,b is illustrated the self-injection process in the case of $B=10T$ in the regimes of single and multiple self-injection. There is no essential difference in the shape of first period of plasma wave (see for example Ref. [7]). However, in the vicinity of maximal density ramps the discontinuity of electron density is broken with clear asymmetry in the axis. With lower magnetic field strength, *B*=6T, the asymmetry is lower as seen in Fig. 7c. We present a zoomed image of that vicinity at the time of self-injection. It is clearly different from the injection process without magnetic field. The transverse structure of the injected electrons is presented in Fig. 7e. The beam shape is not symmetrical as well. The beams are modulated by the magnetic field during their acceleration. This modulation is caused by the drift instability in the crossing electric and magnetic field.

This instability is seen brighter in the case of uniform plasma in the magnetic field as in Fig.8a,b. In uniform plasma the self-injection dominated by relativistic modulation, Larmor modulation, and plasma wave oscillation. It runs longer than the self-injection in plasma channels. However, higher amplitude of pulse waist oscillations results in stronger drift instability: the accelerated electrons are modulated, lower energy electrons are expelled from the acceleration phase of laser wake. In the case of magnetic field the beam loading the beam loading is strong enough to make a plasma cavity as seen in Fig.8b. The transverse beam profile in the cavity has a complicated structure as seen in Fig.8c. The effect of magnetic field in this case cannot be explain by the parametric resonance. We attribute the effect to freezing of lower energy electron background in the external magnetic field. This freezing changes the phase of plasma oscillations. Along with this effect the external magnetic field makes a contra-propagating electron fluxes in the vicinity of the maximal electron density or minimal electron fluid velocity (in the reference frame moving with the pulse group velocity). In higher density cases it may provoke the 'two-beam' instability. However, in some runs we do not observe any effect of magnetic field in high density plasma.

The energy distributions of accelerated electron are given in Fig.9. One can see that there is no notable difference in the energy distribution near the maximal energy in the case $B$=0T and $B$=10T for the acceleration in the plasma channel. However, an essential difference is observed at lower energies. In the uniform plasma, the number of accelerated electron flux through the central 30 mm hole is lower than that is the plasma channel at B=10T; the maximal energy is expectedly smaller. The lower energy electron distribution is similar to that in the plasma channel. However, a small variation of the plasma and focusing laser parameters result in considerable change in the electron energy distribution.

**IV. Conclusion**

We have demonstrated effect of external static magnetic field with strength $B$<100T on the electron-self injection and acceleration by a laser wake field. At lower plasma densities, typical for the capillary discharges used for GeV electron acceleration, in the reference frame moving with the group velocity of laser pulses, the electron Larmor frequency in the external magnetic field becomes comparable with the plasma frequency affecting on the condition of the parametric resonance and making the charge loading less-dependant on the plasma characteristics. This effect improves stability of laser wake field acceleration and allows a good control of charge loading in the acceleration phase of laser wake and, therefore, a control of maximal energy of accelerated electrons. However an increase of the magnetic field strength results in the geometrical emittance of accelerated electrons. At $B$>10 T the electron bunches propagate in a cone with the angle increasing with the magnetic field strength.

At higher plasma density we have also observed the effect of magnetic field resulting in the beam characteristics. We attribute this effect to two effects: to lower energy electron core freezing and to the two beam instability in the vicinity of the maximal density at plasma wave. However in the high density cases the effect of magnetic field may be much stronger at the pre-pulse level where effect of magnetic field on the dynamic of plasma ignited by laser pre-pulses is usually much stronger [3].

**Figure Captions**

**Fig. 1** Dynamics of laser field in the propagation in a plasma channel: (a) *L*=0.1 mm, (b) *L*=0.7 mm (c) *L*=1.5 mm (the pulse depletion is 50%). Arrows show the field of accelerated electrons.

**Fig. 2** Dynamics of laser field in the propagation in uniform plasma: (a) *L*=0.1 mm, (b) *L*=0.7 mm (c) *L*=1.5 mm (the pulse depletion is 70%) Arrows show the field generated by accelerated electrons.

**Fig. 3** Spatial distribution of accelerated electrons in the plasma channel at *B*=0T (a), *B*=10T(b), *B*=20T(c), and *B*=100T(d): includes electrons form two planes *y*=0 and *z*=0. Symbols represent the energy of electron in MeV.

**Fig. 4** Spatial distribution of energetic electrons in the laser wake in the plasma channel for different magnetic field strengths: *B*=0T (a), *B*=10T(b), *B*=20T(c), and *B*=100T(d).

**Fig.5** Dependency of charge of accelerated electron on the energy at different magnetic field strengths.

**Fig. 6** The charge loaded dependency on the plasma parameters as in Ref. [6] with and without magnetic field.

**Fig. 7** Electron density distribution in the wake of a laser pulse in the higher density plasma channel in the presence of magnetic field: (a), (b), (c),(d) $B$=10T, (c) 5T

**Fig. 8** Electron density distribution in the wake of a laser pulse in uniform plasma in the presence of magnetic field $B$=10T.

**Fig. 9** Energy distribution of accelerated electrons out of the plasma: (1),(2) plasma channel, (3) uniform plasma, (1)$B$=0T, (2),(3) $B$=10T

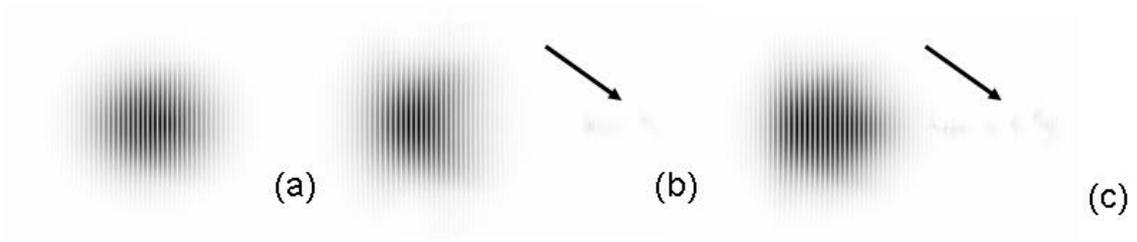

**Fig. 1**

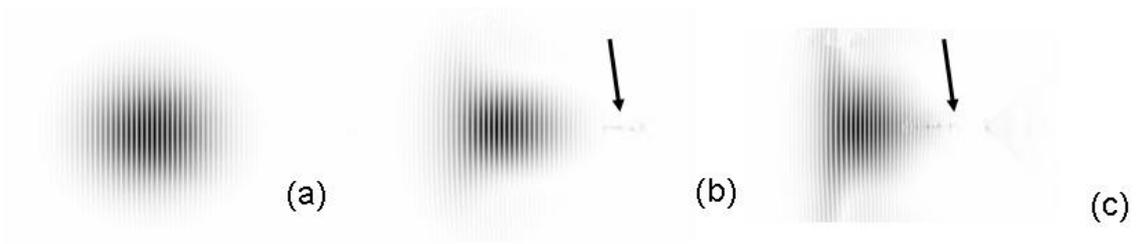

**Fig. 2**

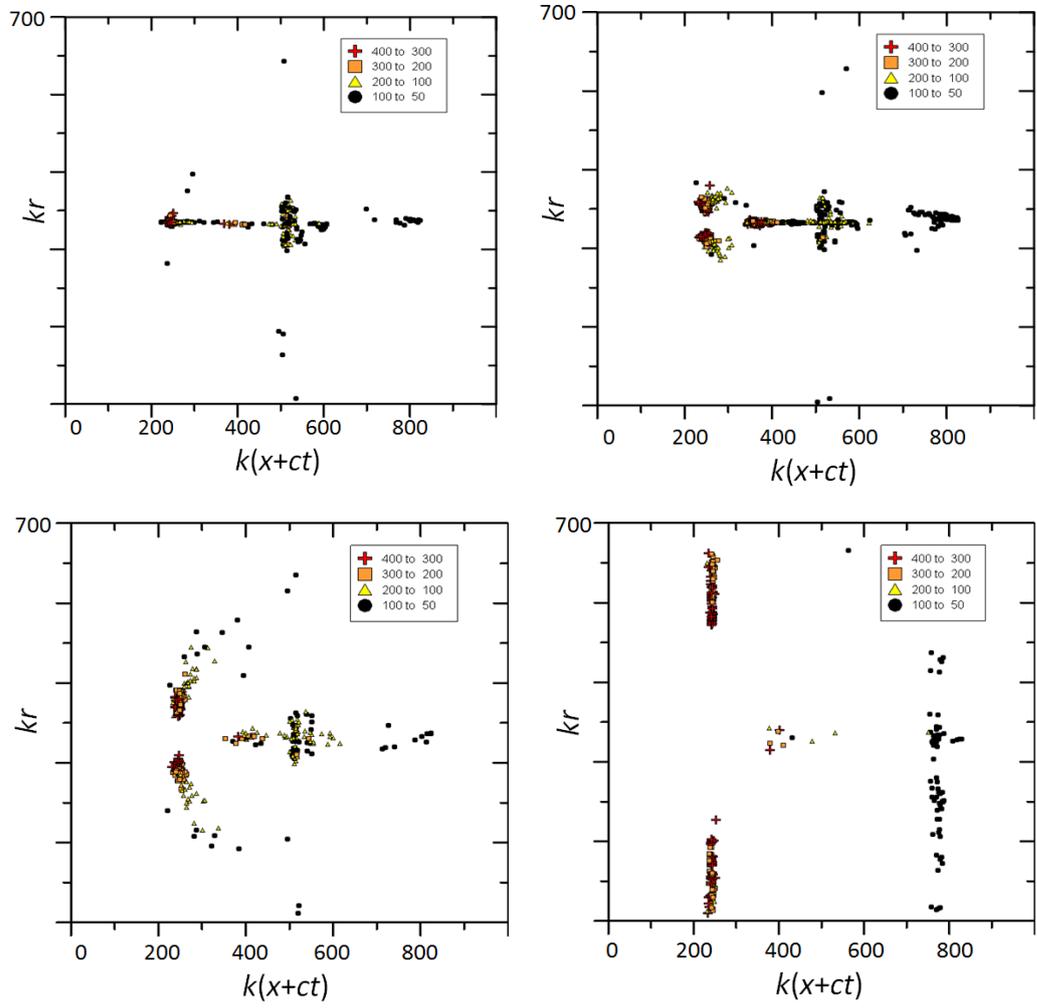

**Fig. 3**

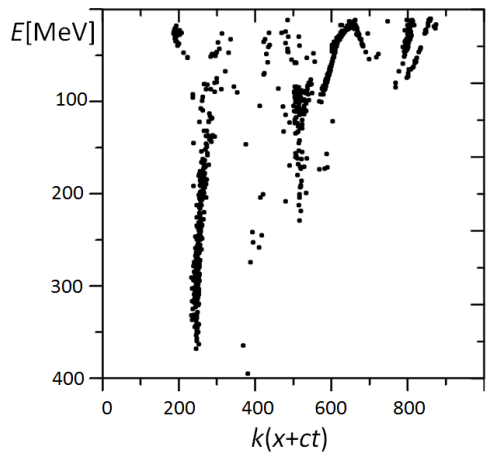
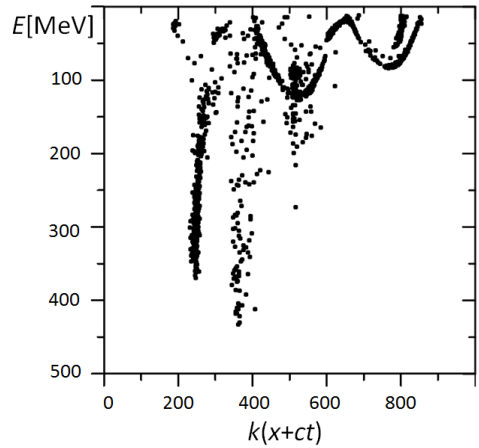
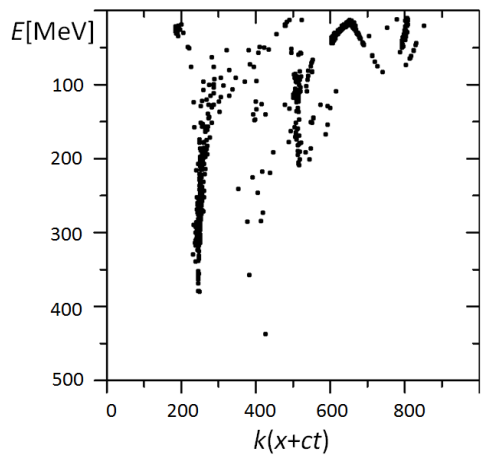
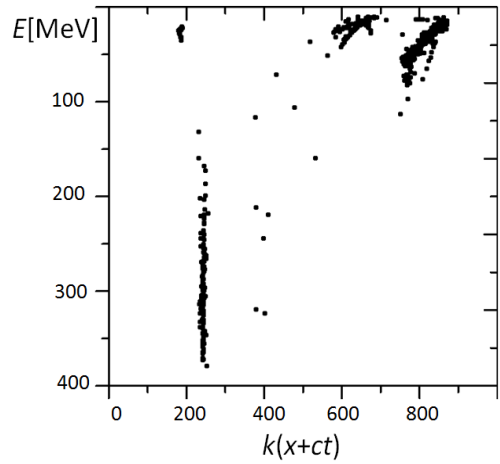

**Fig. 4**

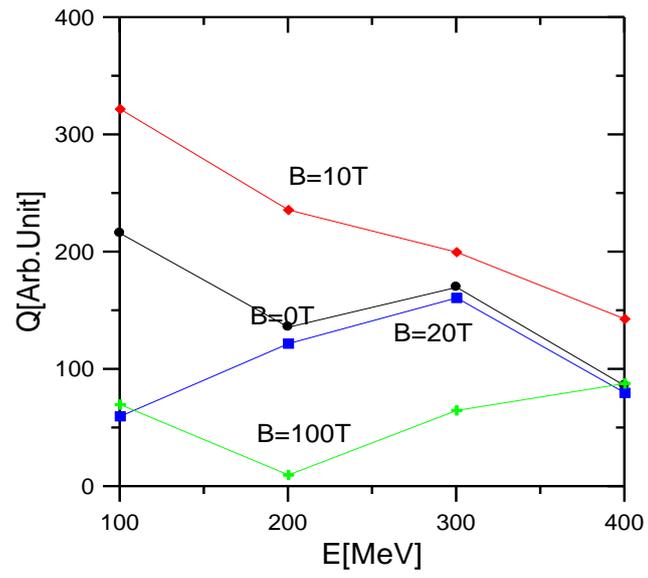

Fig.5

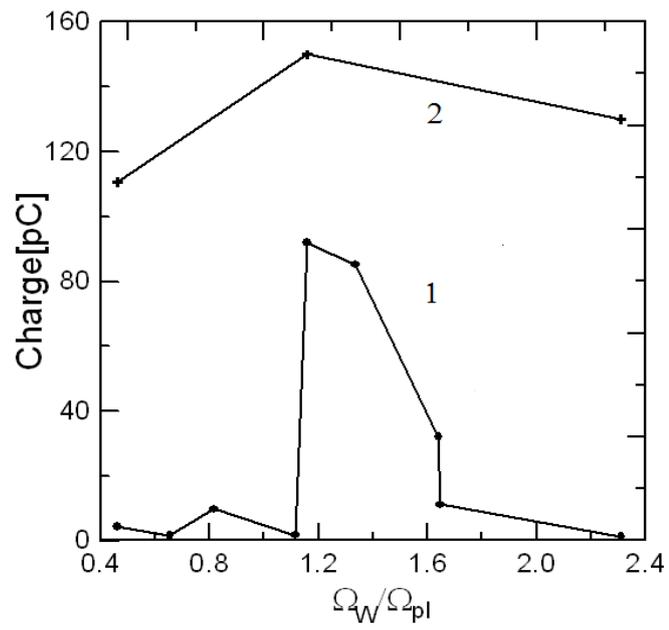

Fig. 6

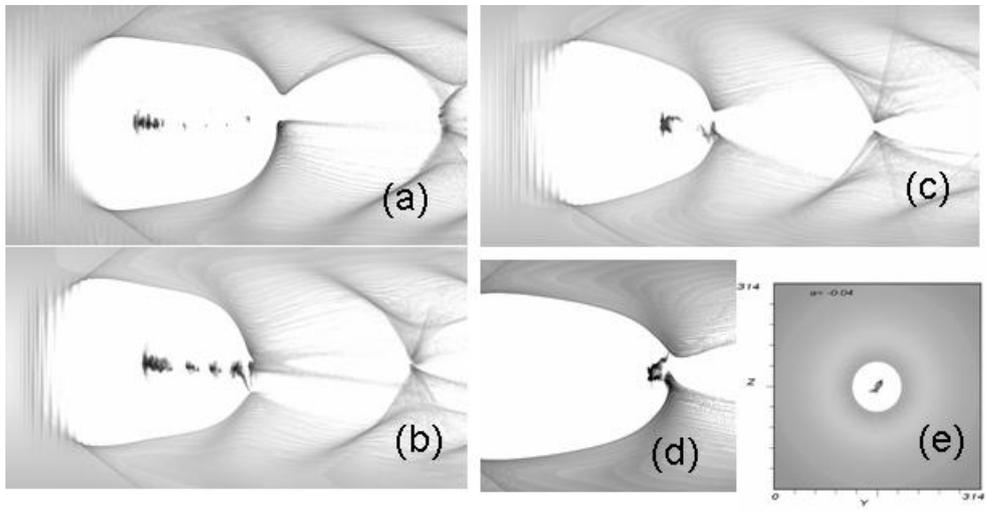

**Fig. 7**

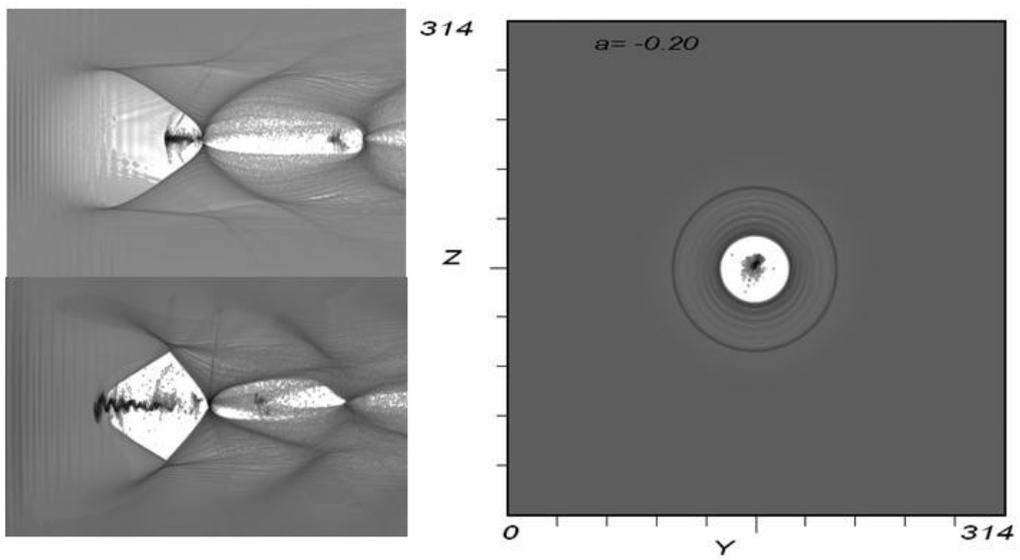

**Fig. 7**

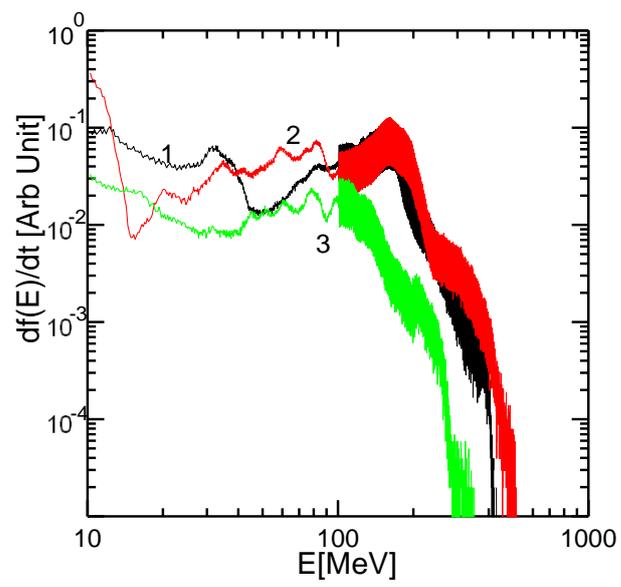

**Fig. 9**